\documentclass[12pt]{iopart}

\usepackage{graphicx}
\usepackage{iopams}  

\begin{document}

\title {Properties of dipolar bosonic quantum gases at finite temperatures}

\author{Abdel\^{a}ali Boudjem\^{a}a }

\address{Department of Physics,  Faculty of Exact Sciences and Informatics, Hassiba Benbouali University of Chlef P.O. Box 151, 02000, Ouled Fares, Chlef, Algeria.}
\ead{a.boudjemaa@univ-chlef.dz}
\vspace{10pt}
\begin{indented}
\item[]February 2014
\end{indented}

\begin{abstract}
The properties of ultracold quantum gases of bosons with dipole-dipole interaction is investigated at finite temperature in the frame of the representative ensembles theory.
Self-consistent coupled equations of motion are derived for the condensate and the non-condensate components.
Corrections due to the dipolar interaction to the condensate depletion, the anomalous density and thermodynamic quantities such as 
the ground state energy, the equation of state, the compressibility and the presure are calculated in the homogeneous case at both zero and finite temperatures. 
Effects of interaction and temperature on the structure factor are also discussed. 
Within the realm of the local density approximation, we generalize our results to the case of a trapped dipolar gas.
\end{abstract}

\noindent{\it Keywords}: Bose-Einstein condensate, dipolar interaction, representative ensembles theory
%\maketitle

\section{Introduction}
Recent progress in the physics of ultra-cold gases have led to the creation of a Bose-Einstein condensate (BEC) with dipole-dipole interaction (DDI) 
and attracted a great deal of interest in theoretical and experimental studies of weakly interacting Bose gases \cite {Baranov, Pfau, Carr, Pupillo2012}. 
What render such systems particularly important is that the atoms interact via a DDI that is both long ranged and anisotropic and it can be also attractive and repulsive. 
By virtue of this interaction, dipolar gases are expected to open fascinating prospects for the observation of novel quantum phases
in ultracold atomic gases \cite {Pupillo2012, Pfau1}.

At zero temperature, there have been a number of theoretical studies on dipolar BEC in particular on the expansion dynamics \cite{Pfau}, the ground state \cite {Santos, Eberlein, santos1, dell},
elementary excitations \cite {bism, lime,boudjGora}, superfluid properties \cite{Tic, Odell}, solitons and soliton-molecule \cite{Tik, santos2, Adhi} and optical lattices \cite{Mul, Gor}.
However, to date,  finite temperature dipolar BECs remain almost entirely unexplored. 
Among the theories that have been used to investigate the behavior of such systems we can quote for instance
mean field Hartree-Fock-Bogoliubov (HFB) theory \cite{Ron, Blak, Hut, Blak1} and perturbation approach \cite {boudj2015}. 

An interesting alternative approach to the finite-temperature Bose gas is provided
by the so-called representative ensembles which has been proposed by Yukalov one decade ago \cite{Yuk, Yuk1, Yuk4}.
The representative ensembles is a self-consistent approach describing Bose-condensed systems.
It guarantees the selfconsistency of all thermodynamic quantities, the validity of
conservation laws, and a gapless spectrum of collective excitations \cite {Yuk, Yuk1, Yuk4}.
It is based on the Bogoliubov shift of field operators, which explicitly breaks the gauge symmetry. The theory is valid for arbitrary
interacting Bose systems, whether equilibrium or nonequilibrium, uniform or  in the presence of any external potentials, and at any temperature.  
Furthermore, the theory has been used successfully to analyze BEC in optical lattices \cite{Yuk7}, BEC in disordered potentials \cite{YYKG, YG} and the
formation of granular state in trapped BECs \cite{YNB}.

In this paper, we extend the representative ensembles approach to a dipolar Bose gas at finite temperature. Within this theory
we go beyond the zero-temperature mean-field approach where the Gross-Pitaevskii equation is solved numerically and analytically (see for review \cite {Baranov, Pfau, Carr, Pupillo2012}).
This finite-temperature analysis allows us to study in detail how the DDI enhances the thermal fluctuations and the thermodynamics of a BEC.

The rest of the paper is organized as follows. In section \ref {RE}, we review the main features of the representative ensembles theory
which constitutes a relevant model to investigate the properties of dipolar Bose gases at finite temperature.
In section \ref {BG}, we derive self consistently equations of motion for the condensed and the thermal cloud in the trapped case.
Accordingly, in the homogeneous case, analytic formulas are obtained for the noncondensed and anomalous densities at both low and high temperatures.
The profile of such densities is plotted in terms of interaction strength and temperature.
In section \ref {StrucF}, we address effects of the interaction strength, the temperature and the angle between the polarization direction and
the relative separation of the dipoles on the behavior of the static structure factor.
Section \ref {Therm} is devoted to highlight the impact of the DDI on some thermodynamic quantities such as the chemical potential, the ground state energy, 
the compressibility and the pressure.
In section \ref {LDA}, we study the behavior of a dipolar Bose gas in a harmonic trap by means of the local density approximation (LDA).
Our conclusions are presented in section \ref{conc}.\\
\textit {Throughout the paper, the system of units is used, where the Planck constant $\hbar=1$ and the Boltzmann constant $k_B =1$.}

\section{Representative Ensemble for Bose-Condensed Systems} 
\label{RE}

The occurrence of the phenomenon of BEC is ensured by the spontaneous breaking of
global gauge symmetry. This latter is achieved by means of the Bogoliubov shift \cite {Bog, Bog1} for the field operators
\begin{equation} \label{FO}
 \hat\psi({\bf r},t)=\phi({\bf r},t) + \hat\psi_1({\bf r},t) \; ,
\end{equation}
in which $\phi({\bf r},t)$ is the condensate wave function and $\hat \psi_1({\bf r},t)$
is the field operator of noncondensed atoms, satisfying the same Bose
commutation relations as $\hat\psi$. 
%Now all operators of physical quantities are
%defined on the Fock space $\cal{F}(\psi_1)$ generated by the field operator
%$\psi_1^\dagger$. It is important to emphasize that the Fock spaces $\cal{F}(\psi)$
%and $\cal{F}(\psi_1)$ are mutually orthogonal [13,19].
These are linearly independent, being orthogonal to each other,
\begin{equation}
\label{ort}
\int \phi^*({\bf r},t) \hat\psi_1({\bf r},t) \; d{\bf r} = 0 \; .
\end{equation}
The operator of noncondensed atoms on average is zero,
\begin{equation} \label{7}
\langle \hat \psi_1({\bf r},t)\rangle \; = \; 0 \; ,
\end{equation}
The condensate function is normalized to the number of condensed atoms
\begin{equation}
\label{Nor}
N_0 = \int |\phi({\bf r},t)|^2 \; d{\bf r} \; .
\end{equation}
while the number of noncondensed atoms is
\begin{equation}
\label{Nor1}
 N_1 \equiv \int \langle \hat \psi_1^\dagger({\bf r},t) \hat\psi_1({\bf r},t)\rangle \; d{\bf r}.
\end{equation}
Thus, the total number of atoms in the system is the sum
\begin{equation}
\label{TN}
N=N_0+N_1.
\end{equation}
The evolution equations for the variables are obtained by the extremization of an effective action \cite {Yuk2, Yuk3, Yuk4}, which
yields the equation for the condensate function
\begin{equation}
\label{EMC}
i\; \frac{\partial}{\partial t} \; \phi({\bf r},t) = \left< {\frac{\delta H}{\delta\phi^*({\bf r},t)}}\right> ,
\end{equation}
and for the field operator of noncondensed atoms
\begin{equation}
\label{EMNC}
i\; \frac{\partial}{\partial t}\; \hat\psi_1({\bf r},t) = \frac{\delta H}{\delta\hat \psi_1^\dagger({\bf r},t)},
\end{equation}
where the grand Hamiltonian is written as 
\begin{equation}
\label{GH}
H \equiv \hat H - \mu_0 \hat N_0 - \mu_1 \hat N_1 - \hat \Lambda,
\end{equation}
in which
\begin{equation}
\label{Mul1}
\hat\Lambda \equiv \int \left [ \lambda({\bf r},t) \hat\psi_1^\dagger({\bf r},t) +\lambda^*({\bf r},t)\hat\psi_1({\bf r},t) \right ]\; d{\bf r} \; ,
\end{equation}
where $\lambda({\bf r},t)$ is a complex function.\\
The Lagrange multipliers $\mu_0$ and $\mu_1$ guarantee the validity of the normalization conditions (\ref{Nor}) and (\ref{Nor1}), while the Lagrange
multipliers $\lambda({\bf r})$ guarantee the conservation condition (\ref{7}).
Evolution equations (\ref{EMC})  and (\ref{EMNC}) are proved to be identical to the Heisenberg equations of motion  \cite{Yuk4}. \\
To determine explicitly the chemical potential, we first keep in mind that in experiments the total number of particles is fixed. 
Then we write the internal energy (\ref{IE}) in the standard way as \cite {Yuk5}
\begin{equation}
\label{IE}
E = \; \langle H \rangle + \;\mu N,
\end{equation}
connecting $E$ with the average of the grand Hamiltonian $\langle H \rangle$ and with
the system chemical potential $\mu$. At the same time, substituting into
equation (\ref{IE}) the grand Hamiltonian (\ref{GH}), and taking into account condition (\ref{Mul1}), we come to the expression
\begin{equation}
\label{Ene}
E = \; \langle H\rangle +\mu_0 N_0 + \mu_1 N_1.
\end{equation}
Comparing equations  (\ref{IE}) and (\ref{Ene}) gives the definition of the system chemical
potential
\begin{equation}
\label{TCP}
\mu \equiv \mu_0 n_0 + \mu_1 n_1,
\end{equation}
expressed through the Lagrange multipliers $\mu_0$ and $\mu_1$ and the
related atomic fractions
$$
n_0 \equiv \frac{N_0}{N} \; , \qquad n_1 \equiv \frac{N_1}{N}.
$$

In equilibrium, the system statistical operator is defined by minimizing the information functional \cite{Yuk2, Yuk4} uniquely
representing the system with the given restrictions. This results in the statistical operator
\begin{equation}
\label{Ro}
\hat\rho = \frac{1}{Z}\; e^{-\beta H}  ; \qquad   Z =  {\rm Tr}\; e^{-\beta H} \; ,
\end{equation}
with being the inverse temperature $\beta\equiv 1/T$.\\
As a consequence, the grand thermodynamic potential is
\begin{equation}
\label{GP}
\Omega = - T\ln {\rm Tr} \; e^{-\beta H} \;.
\end{equation}

The details of the representative ensembles approach for Bose systems have been thoroughly exposed in \cite {Yuk2,Yuk3,Yuk4,Yuk5, Yuk6}.

\section{Bose Gas with dipole-dipole interaction} \label{BG}

Let us now focus our attention to a system of trapped bosons that interact by both a long-range DDI and a contact
interaction characterized by
\begin{equation}\label{Vdd}
V(\mathbf{r}-\mathbf{r'})=g\delta(\mathbf{r}-\mathbf{r'})+V_d({\bf r}-{\bf r'}), 
\end{equation}
where $g=4\pi \hbar^2 a/m$ corresponds to the short-range part of the interaction and is parametrized by the scattering length as $a$. 
The dipole interaction has the form
\begin{equation}\label{dd}
V_d(\mathbf{r}) =\frac{C_{dd}}{4\pi}\frac{1-3\cos^2\theta}{r^3},
\end{equation}
where the coupling constant $C_{dd} $ is $M_0 M^2$ for particles having a permanent magnetic dipole moment $M$ ($M_0$ is the magnetic permeability
in vacuum) and $d^2/\epsilon_0$ for particles having a permanent electric dipole $d$ ($\epsilon_0 $ is the permittivity of vacuum),
$m$ is the particle mass, and $\theta$ is the angle between the relative position of the particles ${\bf r}$ and the direction of the dipole.
The characteristic dipole-dipole distance can be defined as $r_*=m C_{dd}/4\pi \hbar^2$ \cite{boudjGora}. 

The Hamiltonian energy operator of such a system may be written as
\begin{eqnarray}\label{ham}
\hat H &=\int d^3r \, \hat \psi^\dagger ({\bf r}) \left(-\frac{\nabla^2}{2m}+V_{trap}({\bf r})\right)\hat\psi(\mathbf{r}) \nonumber\\
&+\frac{1}{2}\int d^3r\int d^3r^\prime\, \hat\psi^\dagger(\mathbf{r}) \hat\psi^\dagger (\mathbf{r^\prime}) V(\mathbf{r}-\mathbf{r^\prime})\hat\psi(\mathbf{r^\prime}) \hat\psi(\mathbf{r}),
\end{eqnarray}
where $\hat\psi^\dagger$ and $\hat\psi$ denote, respectively the usual creation and annihilation field operators and $V_{trap}$ is the trapping potential.

%For most polar molecules $r_*$ ranges from 10 to $10^4$ \AA. 
%The disorder potential is described by vanishing ensemble averages $\langle U(\mathbf r)\rangle=0$
%and a finite correlation of the form $\langle U(\mathbf r) U(\mathbf r')\rangle=R (\mathbf r,\mathbf r')$.

Before deriving evolution equations for the condensate wave function and noncondensate operator, let us now introduce the following notations:
the local condensate density
\begin{equation}
\label{CD}
\rho_0 ({\bf r},t)  \equiv|\phi({\bf r},t)|^2,
\end{equation}
the density of uncondensed particles is
\begin{equation}
\label{ND}
\rho_1({\bf r})  \equiv\rho_1({\bf r},{\bf r})  \equiv\langle \hat\psi_1^\dagger({\bf r}) \hat\psi_1({\bf r}) \rangle,
\end{equation}
and the diagonal element of the anomalous density matrix,
\begin{equation}
\label{AD}
\sigma_1({\bf r}) \equiv \sigma_1({\bf r},{\bf r})  \equiv \langle \hat\psi_1({\bf r}) \hat\psi_1({\bf r}) \rangle.
\end{equation}
The total local density is
\begin{equation}
\label{TotD}
\rho ({\bf r}) = \rho_0({\bf r}) + \rho_1({\bf r}).
\end{equation}
While  the single-particle density matrix
\begin{equation}
\label{NCD}
\rho_1({\bf r},{\bf r'})  \equiv\langle \hat \psi_1^\dagger({\bf r'}) \hat\psi_1({\bf r}) \rangle,
\end{equation}
the anomalous density matrix
\begin{equation}
\label{AnomD}
\sigma_1({\bf r},{\bf r'})  \equiv \langle \hat\psi_1({\bf r'}) \hat\psi_1({\bf r}) \rangle ,
\end{equation}
and the anomalous triple correlator
\begin{equation}
\label{Trip}
\xi({\bf r},{\bf r'}) \equiv  \langle \hat\psi_1^\dagger({\bf r'}) \hat\psi_1({\bf r'}) \hat\psi_1({\bf r})\rangle.
\end{equation}

Then, equation (\ref {EMC}) yields the evolution equation for the condensate function
\begin{eqnarray} \label{BEC}
i\;\frac{\partial}{\partial t} \phi({\bf r},t) & =\left (-\frac{\nabla^2}{2m} +V_{trap}({\bf r})-\mu_0 \right ) \phi({\bf r},t) \\ \nonumber
&+ g \left [ \rho({\bf r}) \phi({\bf r}) + \rho_1({\bf r})\phi({\bf r}) 
+\sigma_1({\bf r})\phi^*({\bf r}) + \xi({\bf r}) \right ]\\ \nonumber
&+\int d^3r' V_d({\bf r}-{\bf r'})\left [ \rho({\bf r'}) \phi({\bf r})+ \rho_1({\bf r},{\bf r'})\phi({\bf r'}) \right. \\
&\left.+\sigma_1({\bf r},{\bf r'})\phi^*({\bf r'}) +\xi({\bf r},{\bf r'})\right].  \nonumber
\end{eqnarray}
This is a general equation for the condensate wave function in the case of a dipolar Bose system. 
No approximation has been involved in deriving it. \\
The equation of motion for the field operator of noncondensed atoms follows
from equation (\ref {EMNC}) giving
\begin{eqnarray} \label{NBEC}
i\; \frac{\partial}{\partial t}\; \hat\psi_1({\bf r},t) &= \left ( -\; \frac{\nabla^2}{2m} +V_{trap}({\bf r}) - \mu_1\right ) \hat\psi_1({\bf r},t) \\ \nonumber 
&+ g \left [ 2\rho_0({\bf r}) \hat\psi_1({\bf r}) +\phi^2({\bf r}) \hat\psi_1^\dagger({\bf r}) + \hat X({\bf r}) \right ] \; \\ \nonumber
&+\int d^3r' V_d({\bf r}-{\bf r'}) \left [ \rho_0({\bf r'}) \hat\psi_1({\bf r}) +\phi({\bf r})\phi^*({\bf r'}) \hat\psi_1({\bf r'}) \right. \\
&\left.+\phi({\bf r})\phi({\bf r'}) \hat\psi_1^\dagger({\bf r'})+\hat X({\bf r'},{\bf r}) \right ]. \nonumber 
\end{eqnarray}
where the last term is the correlation operator
\begin{eqnarray}
\hat X({\bf r},{\bf r'})& \equiv  \hat\psi_1^\dagger({\bf r'}) \hat\psi_1({\bf r'}) \phi({\bf r}) +\hat\psi_1^\dagger({\bf r'})\hat\psi_1({\bf r}) \phi({\bf r'}) +
 \hat\psi_1({\bf r'}) \hat\psi_1({\bf r}) \phi^*({\bf r'}) \nonumber\\
& +\hat\psi_1^\dagger({\bf r}) \hat\psi_1({\bf r'}) \hat\psi_1({\bf r}) \; \nonumber.
\end{eqnarray}
One can easily check that equations (\ref {BEC}) and (\ref {NBEC}) satisfy all the conservation laws. 
For $\rho_1=\sigma_1=0$, equations (\ref {BEC}) and (\ref {NBEC}) reduce to the standard Gross-Pitaevskii equation with DDI interactions (see for review \cite {Baranov, Pfau, Carr, Pupillo2012}) which describes BEC only at zero temperature. Putting $C_{dd}=0$, one recovers the usual representative ensembles equations for BEC with pure contact interactions \cite{Yuk2, Yuk3}.
Moreover, although the numerical simulation of equations (\ref {BEC}) and (\ref {NBEC}) is somehow difficult due to the nonlocality of the DDI it remains relatively easier
compared to that of the standard HFB equations \cite{Ron, Hut}. The reason is that this latter approach becomes rapidly unstable for higher modes in particular for large number of particles.

In the static case, one has that the condensate function be time independent.
Therefore, setting to zero the right-hand sides of equation (\ref {BEC}), we get 
\begin{eqnarray} \label{BEC1}
\mu_0 \phi({\bf r})  &=\left (-\frac{\nabla^2}{2m} +V_{trap}({\bf r})\right ) \phi({\bf r},t) \\
&+ g \left [ \rho({\bf r}) \phi({\bf r}) + \rho_1({\bf r})\phi({\bf r}) 
+\sigma_1({\bf r})\phi^*({\bf r}) + \xi({\bf r}) \right ] \nonumber\\
&+\int d^3r' V_d({\bf r}-{\bf r'})\left [ \rho({\bf r'}) \phi({\bf r})+ \rho_1({\bf r},{\bf r'})\phi({\bf r'}) \nonumber \right. \\
&\left.+\sigma_1({\bf r},{\bf r'})\phi^*({\bf r'}) +\xi({\bf r},{\bf r'})\right].  \nonumber
\end{eqnarray}

For a uniform Bose system ($V_{trap}({\bf r})=0$),  the field operator of noncondensed particles transforms as
$\hat\psi_1({\bf r})= (1/V) \sum_{\bf k} \hat a_k e^{i \bf k. \bf r}$ with $V$ being the system volume, 
and the interaction potential (\ref {dd}) is given by \cite {Baranov, boudj2015}
\begin{equation} \label{Four}
 \tilde V (\mathbf k)=g[1+\epsilon_{dd} (3\cos^2\theta_k-1)],
\end{equation}
where the vector ${\bf k}$ represents the momentum transfer imparted by the collision, and 
$\epsilon_{dd}=C_{dd}/3g $ is the dimensionless relative strength which describes the interplay between the DDI and short-range interactions.
The grand Hamiltonian takes the form of a sum
\begin{equation}
\label{HamUnif}
H = \sum_{n=0}^4 H^{(n)},
\end{equation}
of five terms, classified according to the number of the operators $\hat a_k$
or $\hat a_k^\dagger$ in the products. The zero-order term does not contain $\hat a_k$,
\begin{equation}
\label{H0}
H^{(0)} = \left ( \frac{1}{2}\;  \tilde V ({\bf k}) \rho_0  - \mu_0 \right ) N_0.
\end{equation}
%Generally, in order to satisfy condition (8), it is necessary and
%sufficient \cite{Yuk5} that the Hamiltonian (14) would not contain the terms
%linear in $\psi_1$ or $a_k$. This can be achieved by choosing the corresponding Lagrange multipliers $\lambda ({\bf r},t)$ in Eq. (9).
Due the orthogonality condition (\ref{ort}), one has $\hat\Lambda=0$ and $H^{(1)}=0$ automatically. \\
The second-order term is 
\begin{eqnarray}
\label{H2}
H^{(2)} &= \sum_{k\neq 0} \left ( \frac{k^2}{2m} +2\rho_0  \tilde V (k) - \mu_1\right ) \hat a_k^\dagger \hat a_k  \\
&+ \frac{1}{2}\rho_0 \sum_{k\neq 0}  \tilde V (k) \left ( \hat a_k^\dagger \hat a_{-k}^\dagger + \hat a_{-k} \hat a_k\right ). \nonumber
\end{eqnarray}
In the third-order term
\begin{equation}
\label{H3}
H^{(3)} = \sqrt{\frac{\rho_0}{V}} \; 
{\sum_{p,q}}'  \tilde V (q) \left ( \hat a_q^\dagger \hat a_{q-p} \hat a_p +\hat a_p^\dagger \hat a_{q-p}^\dagger \hat a_q \right ),
\end{equation}
the prime on the summation symbol implies that ${\bf p}\neq 0$,
${\bf q}\neq 0$, and ${\bf p}-{\bf q}\neq 0$. \\
The fourth-order term is
\begin{equation}
\label{H4}
H^{(4)} = \frac{1}{2V} \; \sum_k {\sum_{p,q}}'  \tilde V (p) \hat a_p^\dagger \hat a_q^\dagger \hat a_{k+p} \hat a_{q-k}\,,
\end{equation}
where the prime means that ${\bf p}\neq 0$, ${\bf q}\neq 0$, ${\bf k}+{\bf p}\neq 0$, and ${\bf k}-{\bf q}\neq 0$.

Applying the Hartree-Fock-Bogolubov approximation to simplify the higher-order terms (\ref{H3}) and (\ref{H4})
and utilizing the definitions
\begin{equation}
\label{D1}
\omega_k \equiv \frac{k^2}{2m} + \rho  \tilde V (0) + \rho_0  \tilde V ({\bf k}) +\frac{1}{V} \sum_{p\neq 0} n_p  \tilde V ({\bf k}+{\bf p}) - \mu_1,
\end{equation}
and
\begin{equation}
\label{D2}
\Delta_k \equiv \rho_0  \tilde V ({\bf k}) +\frac{1}{V} \sum_{p\neq 0} \sigma_p  \tilde V ({\bf k}+{\bf p}),
\end{equation}
where 
\begin{equation}
\label{Ndistrib}
n_k \; \equiv \; \langle \hat a_k^\dagger \hat a_k \rangle,
\end{equation}
and 
\begin{equation}
\label{ANdistrib}
\sigma_k \; \equiv \; \langle \hat a_k \hat a_{-k} \rangle.
\end{equation}
We obtain then
\begin{equation}
\label{ToHamUn}
H = E_{HFB} + \sum_{k\neq 0} \left [ \omega_k \hat a_k^\dagger \hat a_k +\frac{\Delta_k}{2}\left ( \hat a_k^\dagger \hat a_{-k}^\dagger + \hat a_{-k} \hat a_k \right )\right ],
\end{equation}
where the nonoperator term is
\begin{equation} \label{Enrgy}
E_{HFB} =  H^{(0)} \; - \frac{1}{2}\rho_1^2  \tilde V (0) V -\frac{1}{2V} \sum'_{k,p}  \tilde V ({\bf k}+{\bf p}) (n_k n_p+\sigma_k\sigma_p), 
\end{equation}
in which ${\bf k}\neq 0$ and $ {\bf p}\neq 0$.\\
Hamiltonian (\ref {ToHamUn}) can be diagonalized by means of the Bogoliubov canonical
transformation $\hat a_k=u_k \hat b_k+v^*_{-k} \hat b^\dagger_{-k}$:
$$
u_k^2 = \frac{\omega_k+\varepsilon_k}{2\varepsilon_k}, \qquad
v_k^2 = \frac{\omega_k-\varepsilon_k}{2\varepsilon_k},
$$
with the Bogoliubov spectrum
\begin{equation}
\label{disp}
\varepsilon_k =\sqrt{\omega_k^2 -\Delta_k^2}.
\end{equation}
Then one gets
\begin{equation}
\label{DiagH}
H = E_B + \sum_{k\neq 0} \varepsilon_k \hat b_k^\dagger \hat b_k,
\end{equation}
where
\begin{equation}
\label{GroundEgy}
E_B \equiv E_{HFB} + \frac{1}{2}\; \sum_{k\neq 0} (\varepsilon_k -\omega_k).
\end{equation}
Under Hamiltonian (\ref{DiagH}), the grand thermodynamic potential (\ref{GP}) takes the form
\begin{equation}
\label{GP1}
\Omega =E_B+ T \sum_k\ln (1- \; e^{-\beta \varepsilon_k}).
\end{equation}
By the Bogoliubov \cite {Bog2} and Hugenholtz-Pines \cite{HP} theorems, the spectrum is to be gapless, which implies that
\begin{equation}
\label{limit}
\lim_{k\rightarrow 0} \varepsilon_k = 0 \; , \qquad \varepsilon_k\geq 0.
\end{equation}
From here it follows that
\begin{equation}
\label{Chim1}
\mu_1= \rho  \tilde V(0)+ \frac{1}{V}  \sum_k (n_k -\sigma_k)  \tilde V ({\bf k}).
\end{equation}
The condensate multiplier in the HFB approximation, when $\xi=0$, becomes
\begin{equation}
\label{Chim0}
\mu_0= \rho  \tilde V(0)+ \frac{1}{V}  \sum_k (n_k +\sigma_k)  \tilde V ({\bf k}).
\end{equation}
%The leading terms in (\ref{Chim1}) and (\ref{Chim0}) are evaluated in the limit $k\rightarrow 0$ since they account for the condensate,
%and  therefore, are anisotropic in $f({\bf k})$, which is intriguing to the DDI \cite{lime}.\\
The difference between equations (\ref{Chim1}) and (\ref{Chim0})  is given by
\begin{equation}
\label{Ch}
\mu_0-\mu_1=\frac{2}{V}  \sum_k \sigma_k  \tilde V ({\bf k}),
\end{equation}
which means that these chemical potentials are different and they coincide only in the standard Bogoliubov approximation \cite{boudj2015, Bog2} i.e. up to zeroth
order in the perturbation theory.\\
With $\mu_1$ from equation (\ref{Chim1}), expression (\ref{D1}) becomes
\begin{equation}\label{D2}
\omega_k = \frac{k^2}{2m} + \rho_0  \tilde V ({\bf k}) +\frac{1}{V} \sum_{p\neq 0} \left[ n_p  \tilde V ({\bf k}+{\bf p})-n_p  \tilde V ({\bf p})+\sigma_p  \tilde V ({\bf p}) \right].
\end{equation}

It is important to emphasize that the chemical potential $\mu_1$ of equation (\ref{Chim1}) makes the spectrum (\ref{disp}) gapless. 
Furthermore, it is easy to check that the same chemical potential (\ref{Chim1}) follows from the Hugenholtz-Pines relation
$\mu_1 = \Sigma_{11} - \Sigma_{12}$, where $\Sigma_{12}$ and $\Sigma_{11}$ are the normal and anomalous self-energies. They can be written as \cite{Yuk4}
\begin{eqnarray} \label{selfenerg}
\Sigma_{11} (0,0)&= (\rho+\rho_0)  \tilde V(0) +\frac{1}{V} \sum_{p\neq 0} n_p  \tilde V ({\bf p}),\\
\Sigma_{12} (0,0)&=\rho_0   \tilde V (0) +\frac{1}{V} \sum_{p\neq 0} \sigma_p  \tilde V ({\bf p})\nonumber.
\end{eqnarray}
% And the necessity stems from the Bogoliubov theorem \cite {Bog2} which can be formulated as the inequality
%$$
%\vert \mu_1- \frac{k^2}{2m} + \Sigma_{12}({\bf k},0) -\Sigma_{11}({\bf k},0) \vert \leq \frac{k^2}{2m_0} \; ,
%$$

%In the limit $k\rightarrow 0$, the spectrum (\ref{disp}) is of acoustic form $\varepsilon_k=c_k k$, with the sound velocity
%With multiplier (\ref{Chim1}), spectrum (\ref{disp}) takes the form
%\begin{equation}
%\label{Bdisp}
%\varepsilon_k = \sqrt{(ck)^2+\left ( \frac{k^2}{2m}\right )^2 } \; ,
%\end{equation}
%in which the sound velocity is
%\begin{equation}
%\label{sound}
%c_k \equiv \sqrt{\frac{\Delta_k}{m^*}} \; .
%\end{equation}
%in which
%\begin{equation}\label{Delta}
 %\Delta _k =\rho_0 \lim_{k\rightarrow 0} f ({\bf k}) +\frac{1}{V} \sum_{p\neq 0} \sigma_p f({\bf p})\; .
%\end{equation}
 %and the effective mass is \cite{Yuk4}
%\begin{equation}
%\label{mass}
%m^* \equiv \frac{m}{1+\frac{2m}{V} \sum_{p\neq 0} (n_p-\sigma_p) \frac{\partial f({\bf p})}{\partial p^2}}\; .
%\end{equation}

The density of noncondensed particles is 
\begin{equation}
\label{noncd}
\rho_1 = \int n_k \frac{d{\bf k}}{(2\pi)^3}= \int \left[\frac{\omega_k}{2\varepsilon_k} \coth\left(\frac{\varepsilon_k}{2T}\right) -\frac{1}{2}\right] \; \frac{d{\bf k}}{(2\pi)^3},
\end{equation}
while the anomalous average reads
\begin{equation}
\label{anomd}
\sigma_1 =\int \sigma_k\frac{d{\bf k}}{(2\pi)^3}  =-\int \frac{\Delta_k}{2\varepsilon_k} \coth\left(\frac{\varepsilon_k}{2T}\right) \frac{d{\bf k}}{(2\pi)^3} \; .
\end{equation}
The quantity $|\sigma_1|$ can be interpreted as the density of pair-correlated atoms. 
The anomalous density grows with interactions and vanishes in noninteracting systems \cite{boudj2011}.

\section{Quantum and thermal fluctuations} 

In this section, we focus ourselves to the case of asymptotically weak interactions where $g \rightarrow 0$ (i.e. $\sigma_1\rightarrow 0$) 
and $r_*\ll \xi$, with $\xi=\hbar/\sqrt{mng}$ being the healing length.
%calculate the quantum and thermal fluctuations of the condensate.
In the low momenta limit ($k\rightarrow 0$), the spectrum (\ref{disp}) is a sound wave $\varepsilon_k=c (\theta) k$ with
the sound velocity being
\begin{equation}
\label{soundd}
c (\theta)\simeq  c_{\delta}\sqrt{1+\epsilon_{dd} (3\cos^2\theta-1)}\,;\qquad (\sigma_1/\rho_0 \ll 1),
\end{equation}
where $c_{\delta}=\sqrt{g\rho_0 /m}$.\\
Due to the anisotropy of the dipolar interaction, the self energies and the sound velocity acquire a dependence on the propagation direction, which is fixed by the angle
$\theta$ between the propagation direction and the dipolar orientation. This angular dependence of the sound velocity has been confirmed experimentally 
by means of the Bragg spectroscopy analysis\cite {bism}.

For the density of noncondensed atoms, employing (\ref{noncd}), we find at $T=0$
\begin{equation}
\label{dep}
\rho_1 ^{T=0}= \int \frac{\omega_k-\varepsilon_k}{2\varepsilon_k} \; \frac{d{\bf k}}{(2\pi)^3} =\frac{(mc_{\delta})^3}{3\pi^2} {\cal Q}_3(\epsilon_{dd}).
\end{equation}
The contribution of the DDI is expressed by the function ${\cal Q}_3(\epsilon_{dd})$, which is special case $j=3$ of
${\cal Q}_j(\epsilon_{dd})=(1-\epsilon_{dd})^{j/2} {}_2\!F_1\left(-\frac{j}{2},\frac{1}{2};\frac{3}{2};\frac{3\epsilon_{dd}}{\epsilon_{dd}-1}\right)$, where ${}_2\!F_1$ 
is the hypergeometric function. Note that the function ${\cal Q}_3(\epsilon_{dd})$ has the following asymptotic behavior for small $\epsilon_{dd}$
$$1+\frac{3 \epsilon_{dd}^2}{10}-\frac{\epsilon_{dd}^3}{35}+\frac{9 \epsilon_{dd}^4}{280}-\frac{3 \epsilon_{dd}^5}{154}+O\left(\epsilon_{dd}^{11/2}\right).$$
All the functions ${\cal Q}_j(\epsilon_{dd})$ are imaginary for $\epsilon_{dd}>1$.\\
We see from equation (\ref{dep}) that the corrections to the quantum depletion are proportional to $n_c$ instead of $n$ as in the standard Bogoliubov method
used in \cite{lime}. 

For the anomalous average (\ref{anomd}), we have at $T=0$
\begin{equation}
\label{anom}
\sigma_1^{T=0} = -\; \frac{1}{2}\; \int \frac{\Delta_k}{\varepsilon_k}\; \frac{d{\bf k}}{(2\pi)^3} \; .
\end{equation}

The integral $\int d{\bf k}/\varepsilon_k$ in equation (\ref {anomd}) is ultraviolet divergent.
To overcome this, we can resort to the standard procedure of analytic
regularization \cite{Zin, Klein, Ander}. For this purpose, we, first, consider the integral
in the limit of asymptotically small $g$, when the dimensional
regularization is applicable, and then analytically continue the result
to arbitrary interactions. The dimensional regularization gives
$$
\int \frac{1}{\varepsilon_k}\; \frac{d{\bf k}}{(2\pi)^3} = -\frac{2}{\pi^2}\; m^{3/2} \; \sqrt{\rho_0 g}\; {\cal Q}_3(\epsilon_{dd}).
$$
Inserting this in equation (\ref {anomd}), we obtain
\begin{equation}
\label{anom1}
\sigma_1^{T=0} = \frac{(m c_{\delta})^2}{\pi^2}\; \sqrt{m\rho_0 g} \; {\cal Q}_3(\epsilon_{dd}).
\end{equation}
Expressions (\ref {dep}) and (\ref {anom1}) show that the noncondensed and the anomalous densities increase monotocally with $\epsilon_{dd}$. 
For a condensate with pure contact interactions (${\cal Q}_3(\epsilon_{dd}=0)=1$), $\rho_1 ^{T=0}$ and $\sigma_1^{T=0}$ reduce to their usual expressions. 
While, for maximal value of DDI i.e. $\epsilon_{dd}\approx1$, they are 1.3 larger than their values of pure contact interactions 
which means that the DDI may enhance fluctuations of the condensate at zero temperature. 
It has been shown also that in the roton regime, the anomalous density becomes peaked giving rise to strongly enhance the density fluctuations 
in both clean \cite{Blak2}  and disordered Bose gas with DDI \cite{boudj2016}.
%In the limit $\sigma_1/\rho_0\rightarrow 0$, $c\rightarrow c_0$, thus, the condensed depletion (\ref {dep}) and the anomalous fraction (\ref {anom1}) become respectively,
%$\rho_1^{T=0}=[(mc_0)^3/(3\pi^2)] {\cal Q}_3(\epsilon_{dd})$, and $\sigma_1^{T=0}=[(mc_0)^3/\pi^2] {\cal Q}_3(\epsilon_{dd})$.
As is seen, the anomalous average is three times larger than the normal fraction of noncondensed particles: $\sigma_1^{T=0}=3\rho_1^{T=0}$
which is in excellent agreement with our recent  result \cite{boudj2015} obtained from the Bogoliubov-Belaev theory.
This emphasizes that the anomalous density becomes principally important and cannot be omitted in interacting Bose condensed systems.

%It is important to stress that the anomalous density (\ref {anom1}) enjoys the natural limiting property
%\begin{equation}
%\label{pro}
%\sigma_1 \rightarrow 0 \qquad (\rho_0 \rightarrow 0) \; .
%\end{equation}
%The physics of property (\ref {pro})  is evident. The existence of both the condensate
%density $\rho_0$ and the anomalous density $\sigma_1$ is caused by the gauge
%symmetry breaking. Both of them are nonzero as soon as the symmetry is broken,
%while both become zero if the symmetry is restored. Any of these quantities
%could be treated as an order parameter for the broken-symmetry phase. So, both
%these quantities, $\rho_0$ and $\sigma_1$, have to nullify simultaneously, when one of them tends to zero.

We now generalize the above obtained results to the case of a spatially homogeneous dipolar
Bose-condensed gas at finite temperature.  \\
At temperatures $T\ll g \rho_0 $, the main contribution to integrals (\ref{noncd}) and (\ref{anomd}) comes from the region of small momentum where $\varepsilon_k= c_{dd} k$.
After some algebra, we obtain the following expressions for the thermal contribution of the noncondensed and anomalous densities 
which we denote them as  $\rho_1^T$ and $\sigma_1^T$, respectively:
\begin{equation}\label {thfluc}
\rho_1^T =-\sigma_1^T = \frac{(mc_{\delta})^3}{12} \left(\frac{T}{mc_{\delta}^2}\right)^{2}{\cal Q}_{-1}(\epsilon_{dd}), %\left(\frac{n_c}{n}\right)^{-1/2}.
\end{equation}
where ${\cal Q}_{-1}(\epsilon_{dd})$ behaves for small $\epsilon_{dd}$ as
$${\cal Q}_{-1}(\epsilon_{dd})=1+\frac{3 \epsilon_{dd}^2}{10}-\frac{\epsilon_{dd}^3}{7}+\frac{3 \epsilon_{dd}^4}{8}-\frac{9 \epsilon_{dd}^5}{22}+O\left(\epsilon_{dd}^{11/2}\right).$$
%\begin{equation}\label {fluc}
%\tilde{n}_T =|\tilde{m}_T| =\frac{mT^2}{12\hbar^3 c_s} {\cal Q}_{-1}(\epsilon_{dd}).
%\end{equation}
Equation (\ref{thfluc}) shows clearly that $\rho_1^T $ and $\sigma_1^T$ are of the same order of magnitude at low temperature and only their signs are opposite.  \\
The situation is quite different at temperatures $T\gg g \rho_0$, where the main contribution to integrals (\ref{noncd}) and (\ref{anomd}) comes from the single particle excitations. 
Hence, $\rho_1^T\approx (mT/2\pi)^{3/2}  \zeta (3/2)$, where  $\zeta (3/2)$ is the Riemann Zeta function. 
The obtained $\rho_1^T$ is nothing else than the density of noncondensed atoms in an ideal Bose gas. Whereas the anomalous density being proportional  to $-(mc_{\delta}^2/2\pi) T$
in the vicinity of the condensation temperature.\\
The total noncondensed and anomalous densities are given, respectively by $\rho_1=\rho_1^{T=0}+\rho_1^T$ and $\sigma_1=\sigma_1^{T=0}+\sigma_1^T$.\\
For $\epsilon_{dd}\approx1$, thermal fluctuations (\ref{thfluc}) are 10.7 greater than their values of a pure short range interaction.
This reflects that the DDIs may strongly enhance fluctuations of the condensate at finite temperature than at zero temperature. 

\begin{figure} 
\centerline{
\includegraphics{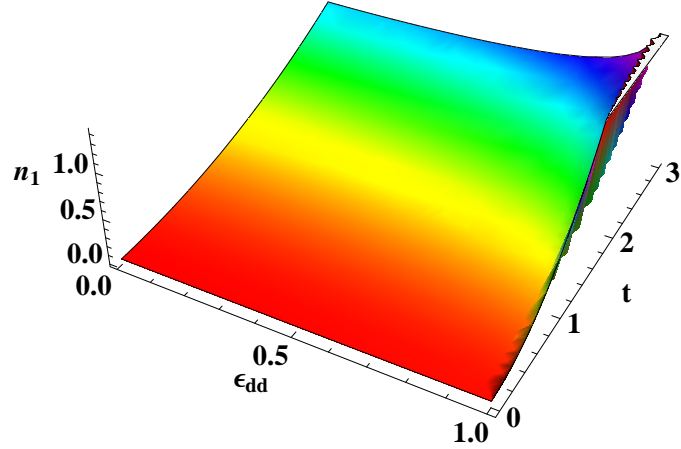}}
 \caption{Noncondensed fraction $n_1$ as a function of the dimensionless temperature $t$ and of the interaction strength $\epsilon_{dd}$.}
\label{Nondens} 
\end{figure}

It is convenient to introduce the following dimensionless quantities: the gas parameter which measures the interaction strength
\begin{equation}
\label{60}
\gamma \equiv \rho^{1/3} a_s,
\end{equation}
the dimensionless temperature
\begin{equation}
\label{61}
t \equiv \frac{mT}{\rho^{2/3}},
\end{equation}
the sound velocity
\begin{equation}
\label{62}
s \equiv \frac{mc_{\delta}}{\rho^{1/3}} \; ,
\end{equation}
the fraction of noncondensed atoms
\begin{equation}
\label{63}
n_1 \equiv \frac{\rho_1}{\rho} \; ,
\end{equation}
and the anomalous fraction
\begin{equation}
\label{64}
\sigma \equiv \frac{\sigma_1}{\rho} \; .
\end{equation}
In terms of these notations, the dimensionless velocity (\ref{62}), satisfies the equation
\begin{equation}
\label{65}
s^2 = 4\pi \gamma n_0 \;,
\end{equation}
where
\begin{equation}
\label{66}
n_0 = 1-n_1,
\end{equation}
stands for the condensed fraction.\\
At sufficiently low temperature, the noncondesend fraction reads
\begin{equation}
\label{67}
n_1 = \frac{s^3}{3\pi^2}{\cal Q}_{3}(\epsilon_{dd})+\frac{t^2}{12 s}{\cal Q}_{-1}(\epsilon_{dd})\;.
\end{equation}
The anomalous fraction (\ref{64}) is
\begin{equation}
\label{68}
\sigma = \frac{2s^2}{\pi^{3/2}}\; \sqrt{\gamma n_0 }{\cal Q}_{3}(\epsilon_{dd})-\frac{t^2}{12 s}{\cal Q}_{-1}(\epsilon_{dd}) \; .
\end{equation}
At higher temperatures when $T\rightarrow T_c$, there is copious evidence that the density of the noncondensed particles becomes the dominant quantity while the anomalous density 
is negligibly small.\\
It is clearly seen from figures (\ref{Nondens}) and (\ref{anom}) that both  $n_1$ and $\sigma$ are monotocally increasing with $\epsilon_{dd}$. 
At low temperatures $t \ll 1$, the anomalous density is larger than the noncondensed density in agreement with the above analytical result.
Near the transition $t \sim t_c$, $n_1$ becomes very important while $\sigma$ goes to zero.

\begin{figure} 
\centerline{
\includegraphics {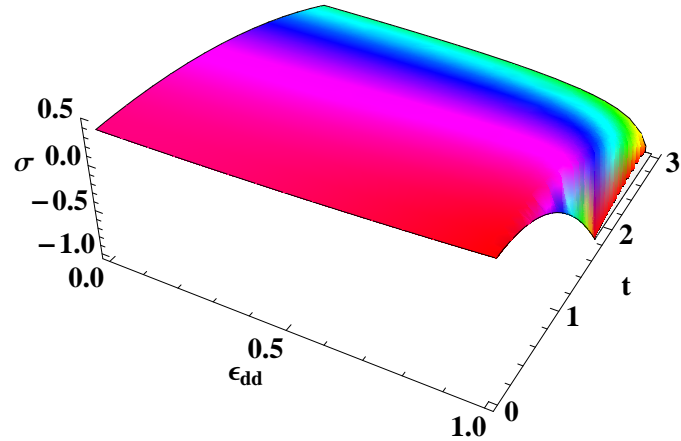}}
 \caption{Anomalous fraction $\sigma$ as a function of the dimensionless temperature $t$ and of the interaction strength $\epsilon_{dd}$.}
\label{anom} 
\end{figure}

\section{Structure factor} \label{StrucF}

The static structure factor which is the Fourier transform of the density-density correlation function is defined through the relation
\begin{equation}   \label{SF}
S({\bf k})= \frac{1}{N}\langle \hat n_k \hat n^\dagger_k\rangle,
\end{equation}
where
$$\hat n_k= \sqrt{N} \left(\hat a_k^\dagger +\hat a_{-k}\right) + \sum_{q\neq 0} \hat a_{k+q}^\dagger \hat a_q.$$
Applying the above Bogoliubov transformation, one obtains for the static structure factor
\begin{equation}   \label{SF1}
S({\bf k})=\frac{E_k}{\varepsilon_k} coth \left(\frac{\varepsilon_k}{2T}\right). 
\end{equation}
The central value $S(0)$ is given as
\begin{equation} \label{SF2}
S(0)=\frac{T}{\mu},
\end{equation}
which is independent from the dipolar force.
For large momenta, the static structure factor approaches unity.

\begin{figure} 
\centerline{
\includegraphics {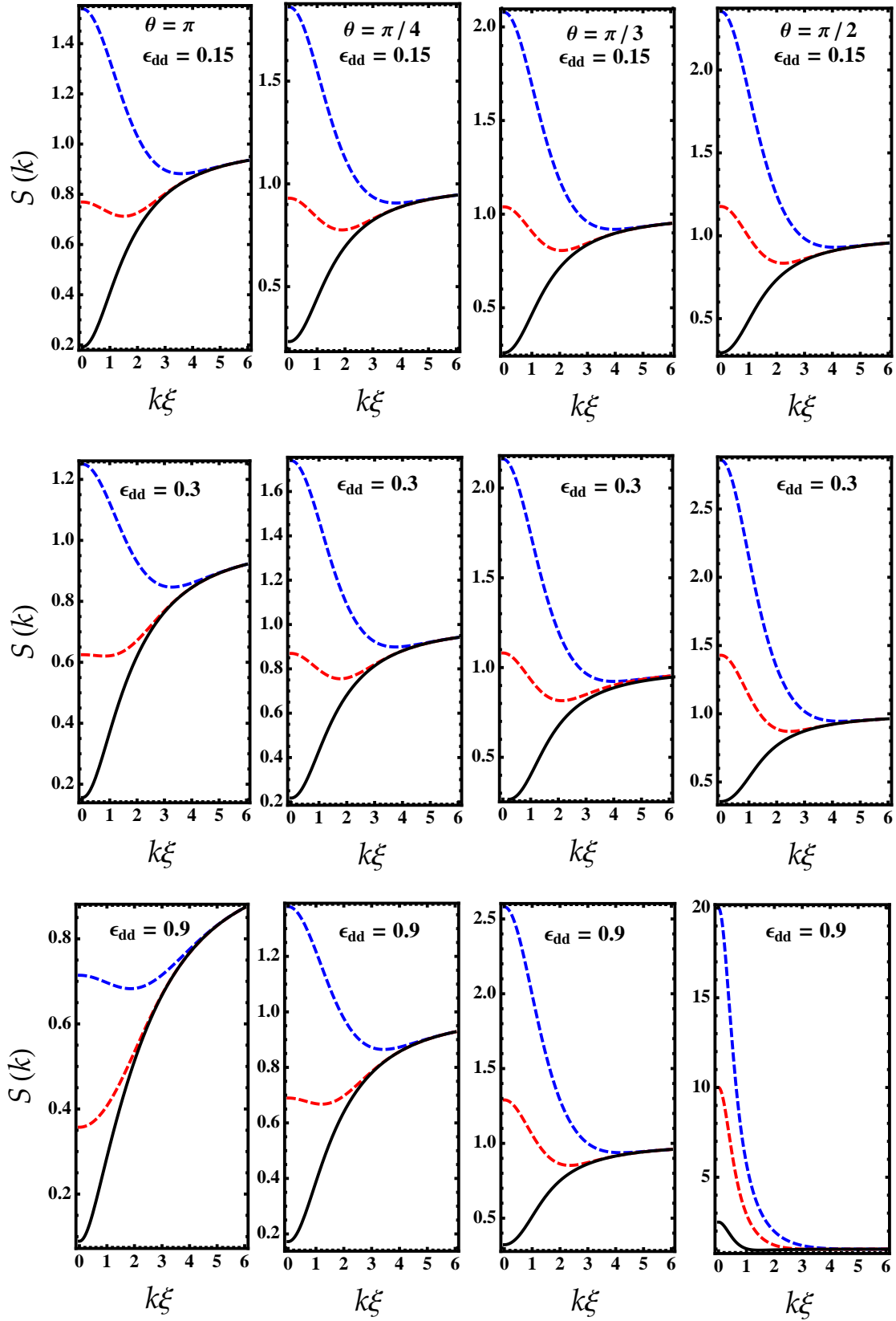}}
 \caption{Static structure factor from Eq.(\ref{SF1}) as a function of dimensionless variable $k\xi$ for different values of temperature, dipolar interaction and the angle $\theta$.
Blue dashed lines: $T/ng=2$. Red dashed lines: $T/ng=1$. Balck solid lines: $T/ng=0.25$.}
\label{SFD} 
\end{figure}

Figure.\ref{SFD} displays the behavior of the static structure factor in terms of the interaction parameter $\epsilon_{dd}$, the temperature $T$ and the angle $\theta$.
We see that $S(k)$ decreases with increasing $\epsilon_{dd}$ and $\theta$. For large DDI  ($\epsilon_{dd} \sim 1$) and $\theta=\pi/2$, $S(k)$ becomes significant
and rises for any value of $T$ indicating that the thermal fluctuations of the density becomes important and therefore, the system may envisage a transition to a new quantum regime.
Note that a similar behavior has been observed in Monte carlo simulations of 2D dipolar Bose gas \cite{Astr, Macia}. 
This effect persists also in the presence of the roton in the excitations spectrum of quasi-2D Bose gas with DDI \cite{Blak3, Blak4}.
In the absence of the DDI, thermal effects are important at small $k$ i.e. in the phonon regime.
%Note that the static structure factor has never been investigated in Bose gas with DDI.

\section{Thermodynamic quantities} \label {Therm}
In this section we examine the effects of the DDIs on the thermodynamics of the system. 
The chemical potential, defined in equation (\ref{TCP}), is expressed through the
Lagrange multipliers (\ref{Chim1}) and  (\ref{Chim0}) and the fractions $n_0$ and $n_1$, which give in the limit $\sigma_1/\rho_0\rightarrow0$
\begin{equation}
\label{ChimP}
\mu =   \tilde V (0)\rho +\sum\limits_{\bf k}  \tilde V ({\bf k})(n_1+\sigma)\rho.
\end{equation}
At $T=0$, the chemical potential turns out to be given as
\begin{equation} \label{ChimPP}
\mu =  g\rho \left[1+\epsilon_{dd} (3\cos^2\theta-1)\right]+\frac{4}{3\pi^2}\; (m c_{\delta})^3\; {\cal Q}_5(\epsilon_{dd}),
\end{equation}
where ${\cal Q}_5(\epsilon_{dd})$ has the following asymptotics for small $\epsilon_{dd}$
$${\cal Q}_5(\epsilon_{dd})=1+\frac{3 \epsilon_{dd}^2}{2}+\frac{\epsilon_{dd}^3}{7}-\frac{3 \epsilon_{dd}^4}{56}+\frac{3 \epsilon_{dd}^5}{154}+O\left(\epsilon_{dd}^{11/2}\right).$$
One can see that for a condensate with a pure contact interaction (${\cal Q}_5(\epsilon_{dd}=0)=1$),
the obtained chemical potential excellently agrees with the famous Lee-Huang-Yang (LHY) quantum corrected equation of state \cite{LHY}. \\
%Noteworthy, that the leading term in Eq. (\ref{ChimPP}) is anisotropic, in the sense that its value depends on the direction
%in which the limit ${\bf k}\rightarrow 0$ is carried out, the subleading term accounts for the quantum fluctuations, and is isotropic.\\
From equation (\ref{Enrgy}), the ground state energy can be written as 
\begin{equation}
\label{FinalEgy}
E_B = E_{HFB} + N \int \frac{\varepsilon_k-\omega_k}{2\rho}\;
\frac{d\bf k}{(2\pi)^3} \; ,
\end{equation}
where the integral is calculated invoking the dimensional regularization \cite{Zin, Klein, Ander}, giving
$$
\int \frac{\varepsilon_k-\omega_k}{2\rho}\; \frac{d\bf k}{(2\pi)^3} =\frac{8m^4c_{\delta}^5}{15\pi^2\rho} \; {\cal Q}_5(\epsilon_{dd})\; .
$$
Under the condition $\sigma_1/\rho_0\rightarrow0$, we obtain for the energy 
%\begin{equation}
%\label{FinalEgy1}
%\frac{E_{HFB}}{N} =\frac{\rho}{2} \sum\limits_{\bf k} f({\bf k}) \left [\left ( n_0^2 - 2n_1^2 -\sigma^2\right) - \mu_0 n_0 \right ]\; .
%\end{equation}
%Summarizing these formulas, we find
\begin{equation} \label{FinalEgy2}
\frac{E}{N} = \frac{\rho}{2} g \left[1+\epsilon_{dd} (3\cos^2\theta-1)\right]+\frac{8m^4c_{\delta}^5}{15\pi^2\rho} \; {\cal Q}_5(\epsilon_{dd}).
\end{equation}
It is important to stress here that our formulas of the equation of state (\ref{ChimP}) and the ground state energy (\ref{FinalEgy2}) constitute a natural extension of those obtained 
from the zeroth order of perturbation theory \cite{lime}. The density $n_c$ of condensed particles is a key parameter instead of the total density $n$.
Note that the ground state energy can be obtained also by integrating the chemical potential correction with respect to the density.

At $T = 0$, the inverse compressibility is equal to $\kappa^{-1} = \rho^2\partial\mu/\partial \rho$. 
Then, using equation (\ref{ChimPP}), we get 
\begin{equation}\label {compr}
\frac{\kappa^{-1}}{\rho^2}=g\left[1+\epsilon_{dd} (3\cos^2\theta-1)\right]+\frac{2}{\pi^2\rho}\; (m c_{\delta})^3\; {\cal Q}_5.
\end{equation}
We see that the chemical potential, the energy and the compressibility are increasing with dipole interaction parameter. 
For $\epsilon_{dd}\approx1$, these quantities are 2.6 larger than their values of pure contact interactions 
which means that DDI effects are more significant for thermodynamic quantities than for the condensate depletion and the anomalous density.

The system pressure can be expressed through the grand potential (\ref{GP1}), which gives
\begin{equation}\label {press}
P =-\frac{\Omega}{V} =-\frac{E_B}{V}-T\int \ln \left(1- \; e^{-\beta \varepsilon_k}\right) \frac{d {\bf k}}{(2\pi)^3}.
\end{equation}
At zero temperature,
\begin{equation}\label {pressT}
P =-\frac{E_B}{V}.
\end{equation}
At temperatures $T\ll g \rho_0$, the thermal pressure can be calculated as
\begin{equation}\label {therpress}
P_T=\frac{\pi^2T^4}{90 c_{\delta}^3} {\cal Q}_{-3}(\epsilon_{dd}).
\end{equation}
where 
$${\cal Q}_{-3}(\epsilon_{dd})=1+\frac{3 \epsilon_{dd}^2}{2}-\epsilon_{dd}^3+\frac{27 \epsilon_{dd}^4}{8}-\frac{9 \epsilon_{dd}^5}{2}+O\left(\epsilon_{dd}^6\right)$$ for small $\epsilon_{dd}$.\\
The inverse isothermal compressibility can be computed easily using $ (\partial P/\partial \rho)_T$.
%\begin{equation}\label {Isocomp}
%\left(\frac{\partial P} {\partial \rho}\right)_T= -\frac{\pi^2T^4}{60 m c_0^5} {\cal Q}_{-3}(\epsilon_{dd}) +\cdots ,
%\end{equation}
%where the zero temperature contribution to the compressibility is given by the expression (\ref{compr}). 

\section{Trapped dipolar BEC} 
\label{LDA}

Here we discuss the case of a harmonically trapped dipolar Bose gas.
To derive analytical expressions for the physical quantities of interest such as the condensate depletion, the equation of state, the ground state energy and so on,
one should use the LDA.  %(for more details on the applicability of this approximation on long-range interactions, see e.g. \cite {lime}).
The LDA or semiclassical approximation is applicable when the external potential is sufficiently smooth, and requires that $u_k({\bf r})$ and $v_k({\bf r})$ 
are slowly varying functions of the position.
Employing  the LDA, the above Bogoliubov equations take then a simple algebraic form as in the homogeneous case. Therefore,
the local Bogoliubov spectrum can be obtained in the usual way and reads $\varepsilon_k({\bf r})=\sqrt{\omega_k^2({\bf r}) -\Delta_k^2({\bf r})}$. 
For sufficiently weak interactions and at low temperature, we obtain for the condensate depletion and the anomalous average, respectively:
\begin{equation}\label {depLADA}
\rho_1({\bf r})=\frac{[mc_{\delta} ({\bf r})]^3}{3\pi^2} {\cal Q}_3(\epsilon_{dd})
+\frac{[mc_{\delta} ({\bf r})]^3}{12} \left(\frac{T}{mc_{\delta} ({\bf r})^2}\right)^{2}{\cal Q}_{-1}(\epsilon_{dd}),
\end{equation}
and 
\begin{equation}\label {anom1LDA}
\sigma_1({\bf r})= \frac{[m c_{\delta} ({\bf r})]^2}{\pi^2}\; \sqrt{m\rho_0({\bf r}) g} \; {\cal Q}_3(\epsilon_{dd}) 
-\frac{[mc_{\delta} ({\bf r})]^3}{12} \left(\frac{T}{mc_{\delta} ({\bf r})^2}\right)^{2}{\cal Q}_{-1}(\epsilon_{dd}).
\end{equation}
On the other hand, the corrections to the energy and the pressure are written in the frame of the LDA, respectively as:
\begin{equation}\label {Reng}
\int \frac{\varepsilon_k-\omega_k}{2\rho}\; \frac{d\bf k}{(2\pi)^3} =\frac{8m^4c_{\delta}^5({\bf r})}{15\pi^2\rho({\bf r})} \; {\cal Q}_5(\epsilon_{dd}),
\end{equation}
and 
\begin{equation}\label {therpress}
P_T=\frac{\pi^2T^4}{90 c_{\delta}^3({\bf r})} {\cal Q}_{-3}(\epsilon_{dd}).
\end{equation}

In the Thomas-Fermi regime where the kinetic term can be neglected from equation (\ref{BEC}), the density $\rho$ in Eqs. (\ref{depLADA}) and (\ref{anom1LDA}) 
can be written in the following algebraic form:
\begin{eqnarray} \label{BECTF}
\rho ({\bf r}) &=\frac{\mu_0 -V_{trap}({\bf r})}{g}- \rho_1({\bf r})-\sigma_1({\bf r}) \\\nonumber
&-\frac{1}{g}\int d^3r' V_d({\bf r}-{\bf r'}) 
\left \{ \rho({\bf r'})+ \frac{1}{\sqrt{\rho_0 ({\bf r})}} \left[\rho_1({\bf r},{\bf r'}) \phi({\bf r'})+\sigma_1({\bf r},{\bf r'}) \phi({\bf r'})\right]\right\}.
\end{eqnarray}
In equation (\ref {BECTF}),  we have omitted the triple anomalous correlator for simplicity.\\
Note that equations (\ref{depLADA}) and (\ref{anom1LDA}) can be  also used to calculate the corrections to the equation of state and the local superfluid density. 
Importantly, we see that the dipolar corrections remain angular independent as in the homogeneous case.
% which is indeed an advantage of the LDA.
%The fact that these corrections remain the same as in the homogeneous case, this an artifact of the LDA. 

\section{Conclusions}
\label{conc}

A self-consistent approach, based on the use of a representative statistical ensemble, developed earlier for Bose-condensed systems, is extended to 
Bose gases with DDI. 
Such an approach allows us to derive a set of coupled equations of motion governing in a self-consistent way the dynamics of the condensate and the thermal cloud.
These equations generalize many of the famous approximations found in
the literature such as the Bogoliubov, the Gross-Pitaevskii and HFB.
Furthermore, we have calculated the corrections to the elementary excitations, noncondensed and anomalous densities
of homogeneous dipolar BEC gases arising from effects of the DDI and temperature.
Moreover, we have found that near the transition temperature, the condensate fraction, anomalous average and the sound velocity tend to zero.
In addition, our results showed that the static structure factor exhibits a strong dependence on the temperature, the interactions strength and the angle $\theta$.
We have pointed out also that the DDI leads to enhance both the quantum and thermal fluctuations as well as the thermodynamic quantities.
Importantly, the shift in these quantities due to the DDI found to be angular independence.
On the other hand, within the LDA, we have extended our results to the case of a harmonically trapped gas.
These results could be directly applied to check how quantum fluctuations can modify the ballistic expansion and collective modes of a trapped dipolar Bose gas.

\section* {Acknowledgements} 
We are grateful to V.I. Yukalov for useful comments and for careful reading of the paper. 
We would like to thank the LPTMS, Paris-sud for a visit, during which part of this work has been done.

\end{document}